# Transition metal single-atom anchored on MoSi$_2$N$_4$ monolayer as highly efficient electrocatalyst for hydrogen evolution reaction

Wei Xun[a*], Xin Liu[a], Qing-Song Jiang[a], Xiao Yang[a*], Yin-Zhong Wu[b], Ping Li[c*]

a Faculty of Electronic Information Engineering, Huaiyin Institute of Technology, Huaian 223003, China

b School of Physical Science and Technology, Suzhou University of Science and Technology, Suzhou 215009, China

c State Key Laboratory for Mechanical Behavior of Materials, Center for Spintronics and Quantum System, School of Materials Science and Engineering, Xi'an Jiaotong University, Xi'an, Shaanxi 710049, China

**Abstract**

Single-atom catalysts are considered as a promising method for efficient energy conversion, owing to their advantages of high atom utilization and low catalyst cost. However, finding a stable two-dimensional structure and high hydrogen evolution reaction (HER) performance is a current research hotspot. Herein, based on the first-principles calculations, we identify the HER properties of six catalysts (TM@MoSi$_2$N$_4$, TM = Sc, Ti, V, Fe, Co, and Ni) comprising transition metal atoms anchored on MoSi$_2$N$_4$ monolayer. The results show that the spin-polarized states appear around the Fermi level after anchoring TM atoms. Therefore, the energy level of the first available unoccupied state for accommodating hydrogen drops, regulating the bonding strength of hydrogen. Thus, the single transition metal atom activates the active site of the MoSi$_2$N$_4$ inert base plane, becoming a quite suitable site for the HER. Based on $\Delta G_{H^*}$, the exchange current density and volcano diagram of the corresponding catalytic system were also calculated. Among them, V@MoSi$_2$N$_4$ ($\Delta G_{H^*}$ = -0.07 eV) and Ni@MoSi$_2$N$_4$ ($\Delta G_{H^*}$ = 0.06 eV) systems show efficient the

HER property. Our study confirms that the transition metal atom anchoring is an effective means to improve the performance of electrocatalysis, and TM@MoSi$_2$N$_4$ has practical application potential as a high efficiency HER electrocatalyst.

## 1. INTRODUCTION

As an emerging and promising energy source, hydrogen has shown its ability to be a substitute for fossil energy [1-7]. Electrocatalytic water splitting to produce hydrogen is a renewable and pollution-free ideal method for Hydrogen evolution reaction (HER) [8, 9]. In order to reduce energy consumption and speed up the reaction kinetics process, catalysts are usually used in the reaction process. To date, the most effective catalysts for the HER activity have been platinum group metals, especially Pt, which are the benchmark for evaluating the activity of other HER catalysts. [10-17]. However, the high price and insufficient resources of platinum group metals as the HER catalysts restrict their wide application. Since the discovery of two-dimensional (2D) graphene in 2004, 2D nanomaterials have become a research focus of catalytic systems due to their unique layered structure and electronic properties. 2D materials such as transition metal sulfide [18-24], MXenes [25-29] and graphene [30-33] improve HER efficiency due to their large specific surface area and more catalytic active sites. Therefore, the development of highly efficient HER catalysts that can be applied in practice has become the main goal of research.

Single-atom catalysts (SACs) have the advantages of high utilization efficiency and clarity of the active site, which provide a boost to the design of catalysts. All kinds of transition metal (TM) atoms supported by two-dimensional materials showed excellent catalytic performance in the catalytic reaction. For example, Pt, Ir, Ru, Pd, Fe, and Ni atoms dispersed on graphene [34-37], and Co, Ni, Pt or Ru atoms anchored on MoS$_2$ [38-40], C$_3$N$_4$ monolayers [41,42], as well as MXenes [43], all show great HER properties. However, the above materials may oxidize or hydrolyze in an unprotected environment (water, oxygen, etc.) [44-46], so the search for more stable

2D materials is still the focus of catalyst design research.

Recently, a new seven-layer material, $MoSi_2N_4$, has been prepared with excellent stability in the natural environment [47]. The electronic and catalytic properties of $MoSi_2N_4$ have been extensively studied [48-56], such as improving HER reaction performance of $MoSi_2N_4$ through doping, tailoring, defect and heterojunctions [57-64]. More crucially, these systems require cumbersome synthesis processes and sophisticated equipment that are difficult to implement in experiments.

In this work, based on first principles calculation, the feasibility of TM atoms modified $MoSi_2N_4$ for electrochemical HER reaction was discussed. We have identified six catalysts with potential HER applications, namely TM atoms anchored to the surface of $MoSi_2N_4$ (represented by TM@$MoSi_2N_4$, where TM = Sc, Ti, V, Fe, Co, and Ni), Among them, V@$MoSi_2N_4$ ($\Delta G_{H*}$ = -0.07 eV) and Ni@$MoSi_2N_4$ ($\Delta G_{H*}$ = 0.06 eV) have the best HER activity. The SACs composed of the TM@$MoSi_2N_4$ system regulate and optimise the empty and occupied states of the *d*-orbitals, making them favorable for the adsorption of $H^+$ and the desorption of $H_2$, thereby improving the efficiency of the HER process. These findings can provide new research ideas for the application of MoSi2N4 materials in the field of catalysis, and identify six kinds of HER catalysts with excellent catalytic performance.

## 2. COMPUTATIONAL METHODS

A 2 × 2 × 1 supercell $MoSi_2N_4$ system was established, and then the SACs were simulated by depositing a transition metal atom (Figures 1a and b). A vacuum region of 20 Å in the z-direction is sufficient to avoid interactions between periodic images. Spin polarization is considered through the Vienna Ab initio Simulation Package (VASP) software [65, 66]. Throughout the calculations, the BEEF-vdw is used to correct for the ubiquitous van der Waals interaction in 2D materials. [67]. The dipole correction is considered for all asymmetric structures. The exchange interaction between electrons is described by the PBE function in the GGA framework. [68, 69]. The cutoff energy of the electron wave function is set as 500 eV, and the convergence standards of energy and force are set as $10^{-6}$ eV and 0.002 eV/Å, respectively. The calculation of K-point sampling for geometric optimization of all systems is 6 × 6 × 1, and 12 × 2 × 1 for the electronic properties. With a Nose-Hoover thermostat at 300 K for 5 ps, We use ab initio molecular dynamics (AIMD) simulations to verify the thermodynamic stability of the corresponding systems [70]. The solvent effect is considered with the implicit solvent model implemented in VASPsol [71].

*Calculating details for HER*. The Gibbs free energy ($\Delta G_{H*}$) is a descriptor proposed by Nørskov and collaborators to describe HER processes. [72]. Under standard conditions, the $\Delta G_{H*}$ of H atom adsorption is calculated by the following formula:

$$\Delta G_{H*} = \Delta E_{H*} + \Delta E_{ZPE} - T\Delta S_{H*} \qquad (1)$$

Where $\Delta E_{H*}$ is the adsorption energy of H, and * is the adsorption site on the surface of $MoSi_2N_4$. At 298.15 K, $\Delta E_{ZPE}$ is the change of zero point energy of adsorption H and gas phase $H_2$, and the corresponding entropy change is $\Delta S_{H*}$.

*Volcano Curve*. At pH = 0, based on the theoretical method proposed by Nørskov and the average $\Delta G_{H*}$ [73], the theoretical exchange current density ($i_0$) of the system is calculated by the following formula:

$$i_0 = -ek_0 \frac{1}{1+\exp\left(\frac{|-\Delta G_{H*}|}{k_b T}\right)} \qquad (2)$$

Where k0 represents the reaction rate constant, and its value is set to 1 when the overpotential is 0. kb is expressed as Boltzmann's constant.

## 3. RESULTS AND DISCUSSION

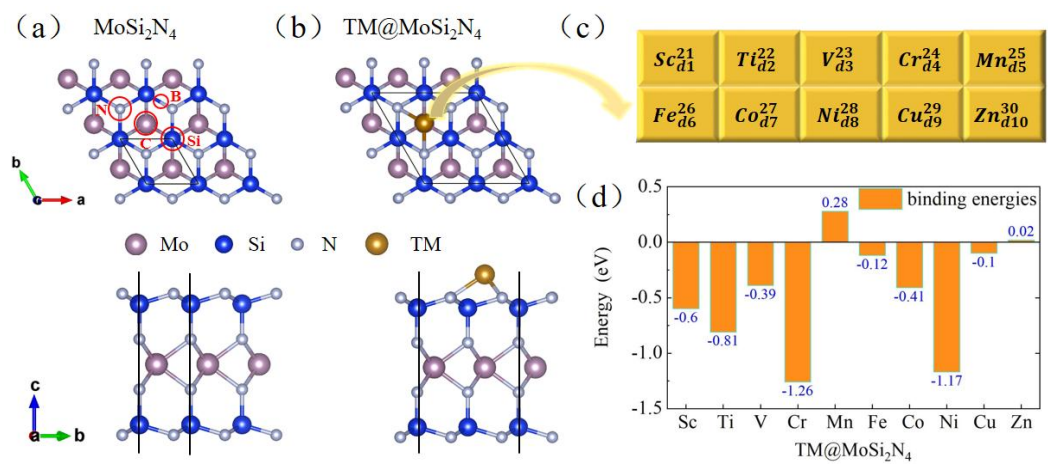

Figure 1 (a) Top and side views of the optimized MoSi$_2$N$_4$ monolayer and (b) surface anchored TM atoms. Red circles denote selected adsorption sites: N, B, Si, and C are denoted as the top of the N atom, the top of the SI-N bond, the top of the Si atom, and the center of the six-membered ring, respectively. (c) 10 transition metal atoms. (d) The binding energy E$_b$ of TM atom anchored into MoSi$_2$N$_4$.

Compared to conventional metal catalysts, SACs typically exhibit higher catalytic performance due to their higher low coordination ratio. [74, 75]. However, TM atoms should be screened because only SACs with a good balance of empty/occupied *d* orbitals exhibit the best catalytic properties. In order to find suitable SACs, ten TM atoms (3d) were selected and chemically adsorbed on the surface of MoSi$_2$N$_4$ (see Figure 1c) and the configuration with the lowest total energy is determined (see supplementary Figure S2). After a comprehensive study of the most favorable energy adsorption configuration (Figure 1a, 1b) and its stability, we preliminarily selected 8 TM atoms (Sc, Ti, V, Cr, Fe, Co, Ni and Cu) as HER catalyst candidates. As shown in Table 1 and Figure S2, the binding energy of Cr and Cu atoms is positive and cannot be stably anchored to the monolayer surface of MoSi$_2$N$_4$.

Table 1 Parameters for the pure TM@MoSi$_2$N$_4$ and H adsorbed TM@MoSi$_2$N$_4$: adsorption site (S$_{ad}$); binding energies of TM atom (E$_{b-TM}$ in eV/atom); charge lost from the adsorbed TM atoms (Q$_{TM}$ in e$^-$/atom); average TM-N bond length (l$_{TM-N}$ in

Å); hydrogen absorption energy ($E_{H^*}$ in eV/molecule) ; TM-H* bond length ($l_{TM-H^*}$ in Å); charge gained by the adsorbed H* ($Q_{H^*}$ in e⁻/atom) and Gibbs free energy of adsorption of hydrogen ($\Delta G_{H^*}$ in eV).

| Catalysts | $S_{ad}$ | $E_{b-TM}$ | $Q_{TM}$ | $l_{Tm-N}$ | $E_{b-H^*}$ | $l_{Tm-H^*}$ | $Q_{H^*}$ | $\Delta G_H$ |
|---|---|---|---|---|---|---|---|---|
| Sc@MoSi$_2$N$_4$ | C | -0.60 | 1.45 | 2.51 | -2.33 | 1.83 | 0.751 | -0.19 |
| Ti@MoSi$_2$N$_4$ | C | -0.81 | 1.38 | 2.43 | -2.37 | 1.76 | 0.776 | -0.24 |
| V@MoSi$_2$N$_4$ | C | -0.39 | 0.41 | 2.49 | -2.21 | 1.62 | 0.807 | -0.07 |
| Fe@MoSi$_2$N$_4$ | C | -0.12 | 0.74 | 2.08 | -2.65 | 1.57 | 0.880 | -0.51 |
| Co@MoSi$_2$N$_4$ | C | -0.41 | 0.50 | 2.07 | -2.56 | 1.54 | 0.690 | -0.42 |
| Ni@MoSi$_2$N$_4$ | C | -1.17 | 0.46 | 2.06 | -2.07 | 1.46 | 0.227 | 0.06 |

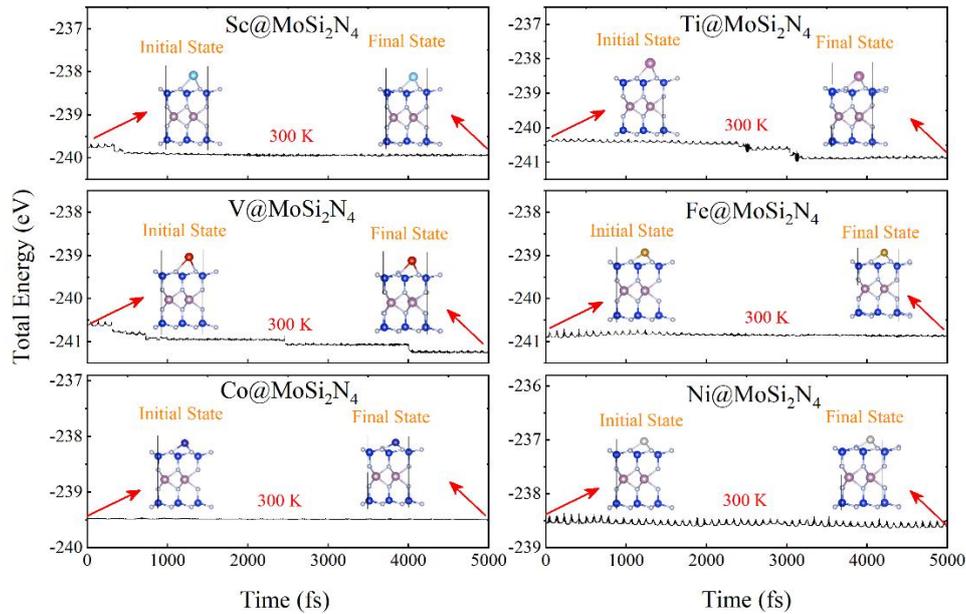

Figure 2 The total energy change with time for TM@MoSi$_2$N$_4$ (TM = Sc, Ti, V, Fe, Co and Ni) at 300 K. The inserts are the initial and final state structures.

To further investigate the structural stability of the selected SACs, we also performed AIMD simulations at TM@MoSi$_2$N$_4$ with a Nose−Hoover thermostat at 300 K. As shown in Figure 2, for Sc, Ti, V, Fe, Co, and Ni, the total energy of the systems fluctuates slightly throughout the simulation and, most importantly, the corresponding crystal structure shows almost no distortion, indicating that these systems are thermodynamically stable. However, the structures of the anchored Cr and Cu atoms have large deformations, suggesting that they are no stable anchored to the surface (see figure S3). Considering the results of binding energy and AIMD, only six TM@MoSi$_2$N$_4$ (TM=Sc, Ti, V, Fe, Co, and Ni) systems will be further investigated.

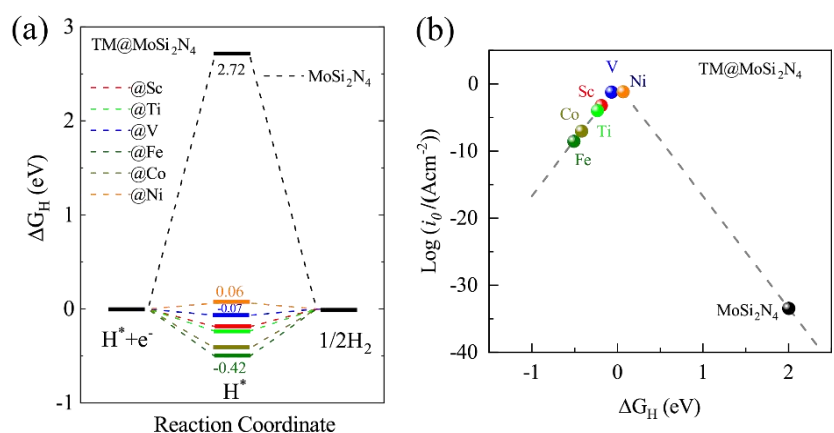

Figure 3 (a) The $\Delta G_{H^*}$ for TM@MoSi$_2$N$_4$ (TM=Sc, Ti, V, Fe, Co, and Ni); (b) the theoretical exchange current $i_0$ of the TM@MoSi$_2$N$_4$.

The catalytic activity of HER in TM@MoSi$_2$N$_4$ system can be described by $\Delta G_{H^*}$. When $\Delta G_{H^*}$ value is zero, the catalytic activity of HER reaches the best. Positive values indicate that the catalyst surface is thermodynamically unfavorable to $H^+$ adsorption, and negative values indicate that it is unfavorable to $H^*$ desorption. For comparison, $\Delta G_{H^*}$ of the initial MoSi$_2$N$_4$ is also calculated separately, at 3.93 eV (Hollow site), 2.83 eV (Si site), and 2.71 eV (N site) (see Figure S1). It indicated that pure MoSi$_2$N$_4$ system is not conducive to $H^+$ adsorption, showing an inert base plane, and is not directly used as HER catalyst. The $\Delta G_{H^*}$ values of TM@MoSi$_2$N$_4$ systems are shown in Figure 3a and summarized in Table 1. Among the systems we studied, V@MoSi2N4 and Ni@TM@MoSi2N4 systems have the lowest ΔGH* values (-0.07eV and 0.06eV, respectively), which are close to 0.00 eV and even better than the well-known Pt (-0.09eV) [57]. Therefore, TM@MoSi2N4 can be used as HER electrocatalyst due to its low cost, feasibility of experimental synthesis and excellent catalytic performance.

In addition, at the equilibrium potential, the inherent rate of electrons in HER process can be reflected by the calculated exchange current density $i_0$, and the peak of the volcano indicates that the system has the best HER performance. [57, 59]. Clearly, the initial MoSi$_2$N$_4$ catalyst with a positive $\Delta G_{H^*}$ value is located in the lower right corner of the volcano curve, while the TM@MoSi2N4 is mostly located near volcanic peaks, which is consistent with the above analysis results. In general, catalysts with excellent HER catalytic properties should satisfy the standard requirement of $|\Delta G_{H^*}| < 0.20$ eV [76, 77]. According to these conditions, V@MoSi$_2$N$_4$ and Ni@MoSi$_2$N$_4$ were successfully screened as potential catalysts for HER.

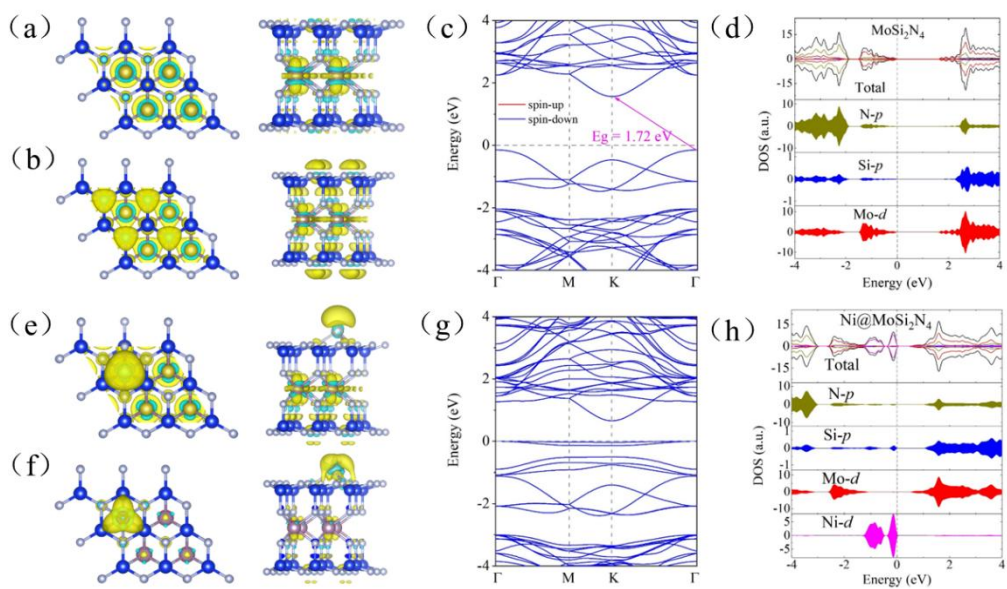

Figure 4 Fukui function (a) F$^+$(r) and (b) F$^-$(r) for MoSi$_2$N$_4$ from top and side views; Charge accumulation and depletion are marked by the yellow and blue regions, respectively. Isosurface value is set to 0.003 e/Å$^3$. (c) Band structure and (d) partial density of states of MoSi$_2$N$_4$. Red and blue lines in (c) correspond to spin-up and spin-down states, respectively. The Fermi level is set to 0 eV. (e) F$^+$(r) and (f) F$^-$(r) for Ni@MoSi$_2$N$_4$, (g) Band structure and (h) partial density of states of Ni@MoSi$_2$N$_4$.

In order to further analyze the HER catalytic activity of TM@MoSi$_2$N$_4$, we studied their electronic structures. As discussed above, at the most stable C sites for adsorption, as shown in Table 1, the strong binding energy, as well as the TM-N bond, indicates that the 6 TM atoms interact strongly enough with the base plane. The charge accumulation between the TM atom and the surrounding N atom (see Table 1 and Figure 5) indicates the strong interaction of the TM-N bond, and is mainly due to the strong hybridization of the N $p$ and TM $d$ orbitals (see Figure 4h and Figure S6). In Table 1, we summarize the average bond length, binding energy and electron transfer of TM-N for the six catalysts, and find that the average bond length of TM-N decreases and the electron transfer decreases with the increase of the metal atomic number. The equilibrium between the vacant orbital and occupied $d$ orbital caused by electron transfer has a significant effect on the catalytic activity of TM@MoSi$_2$N$_4$.

To discuss the activity of different sites in all systems, the Fukui function is calculated. It can be defined by the following equation [57]:

$$F^+(r) = n_{N+1}(r) - n_N(r) \tag{3}$$

$$F^-(r) = n_N(r) - n_{N-1}(r) \tag{4}$$

For systems with N-1, N, and N+1 electrons, the charge densities are denoted by $n_{N-1}$, $n_N$, and $n_{N+1}$, respectively. All use the same crystal structure to ensure the same external potential. For the initial MoSi$_2$N$_4$, as shown in Figure 4a and 4b, both Fukui functions mainly surround the outermost N atom, indicating that H$^+$ is most inclined to be adsorbed on the N atom, which is consistent with the above calculation that $\Delta G_{H^*}$ values are lowest at the N site. Because of the large band gap of the MoSi$_2$N$_4$ monolayer (1.72 eV, see Figure 4c), the first available state occupied by the hydrogen 1$s$ electron is the lowest conduction state, and the energy required to fill the hydrogen 1$s$ electron is large. This relatively weak binding of hydrogen atoms suggests that the initial MoSi$_2$N$_4$ had inertial HER activity.

Unlike the case of MoSi$_2$N$_4$, the Fukui functions (Figures 4e, 4f, and S4) of TM@MoSi$_2$N$_4$ are highly localized around the anchored TM atom, providing the optimal active site for atomic hydrogen adsorption. Figure 4g shows the band structure of Ni@MoSi$_2$N$_4$. It favors a non-magnetic ground state different from the initial MoSi$_2$N$_4$. Compared with the PDOS of the initial MoSi$_2$N$_4$, as shown in Figure 4h, the energy bands across the Fermi level are mainly contributed by the Ni $d$ orbitals. The empty band near the Fermi level is a key factor in the improvement of catalytic activity of HER, as it provides an occupied orbital for the adsorption of H atom and enhances the binding energy between the adsorption site and H atom. As shown in Figures S5 and S6, our calculations show that TM@MoSi$_2$N$_4$ (TM =Sc, Ti, V, Fe, and Co) exhibits spin-polarized ground states. What's more, Fe@MoSi$_2$N$_4$ and Co@MoSi$_2$N$_4$ show semi-metallic features in which one of the spin channels passes through the Fermi level. The large number of electronic states at the Fermi level indicates that the system has a good electrical conductivity and facilitates electron transfer in the HER process. We can see that all systems exhibit metal properties, and the electronic states near the Fermi level are mainly due to the contribution of TM $d$ orbitals. While improving the conductivity and promoting the electron transfer. It single-electron occupied state is conducive to the adsorption of hydrogen, which is very beneficial to the HER.

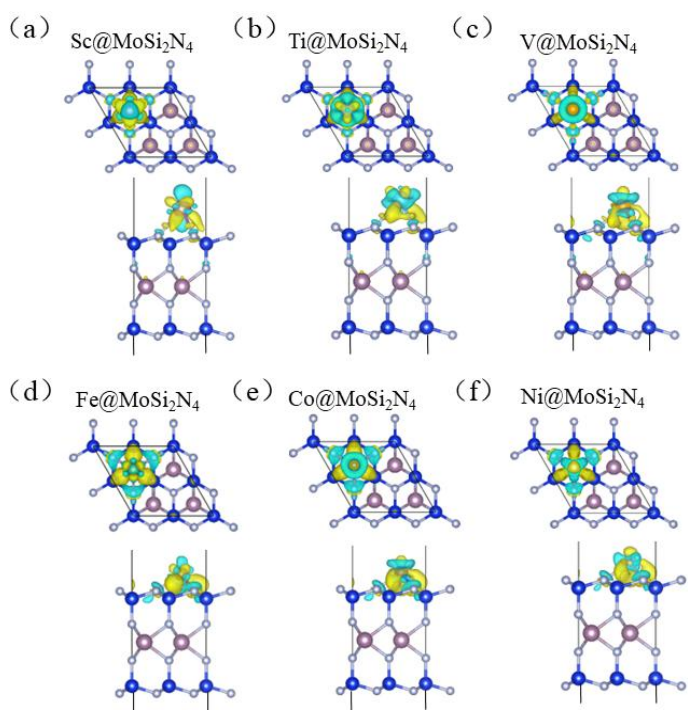

Figure 5 Top and side views of the charge density difference for TM@MoSi$_2$N$_4$ (TM = Sc, Ti, Fe, V, Co and Ni), which are obtained with these optimized configurations. The isosurface is 0.005e e/Å$^3$. Charge accumulation and depletion are represented by yellow and blue areas, respectively.

Next, we calculate the charge density difference for all TM@MoSi$_2$N$_4$ systems,

as shown in Figure 5. Obviously, after anchoring the TM atom, there is an obvious charge redistribution process between the TM and the adjacent N atom. We found that the electron acceptance region on the surface of all systems is predominantly distributed around the TM atom rather than the N atom, making positively charged $H^+$ adsorption feasible and thus providing the optimal active site, which is consistent with the results discussed earlier via the Fukui function. More importantly, we also calculate the Bader charge in order to analyze the charge transfer process quantitatively. Bader charge analysis (Table 1) shows that charge transfer from TM to adjacent N atoms results in unoccupied states near the Fermi level. The Bader charge after hydrogen adsorption is also calculated, as shown in Table 1. The transfer of charge from TM to H atom enhances the adsorption strength of H and HER activity.

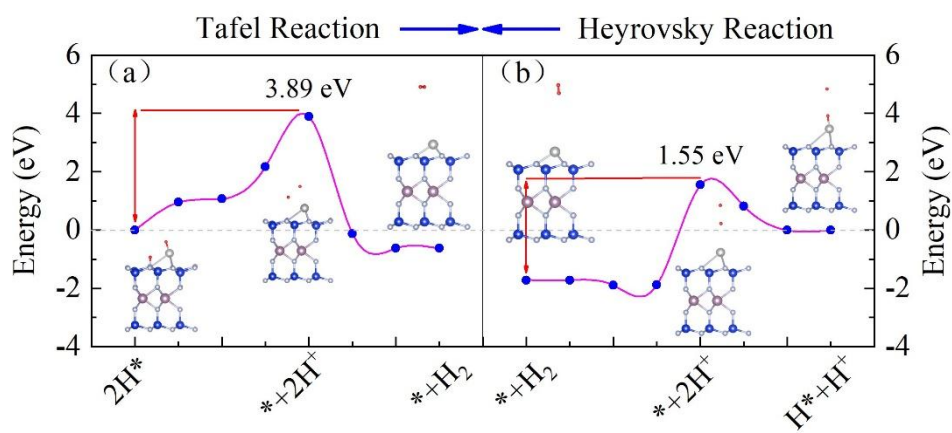

Figure 6 (a) The Volmer-Tafel and (b) Volmer-Heyrovsky reaction pathways of Ni@MoSi$_2$N$_4$ in HER process.

There are Volmer-Heyrovsky (V-H) mechanism and Volmer-Tafel (V-T) mechanism in HER catalytic reaction process [59], and the Volmer step is the same in both mechanisms. In the V-T mechanism, transferred electrons first bind to protons and are adsorbed by catalysts, forming adsorbed hydrogen $H^*$ at the active site of the catalyst (Volmer step); and then two $H^*$ combine to form $H_2$ gas (Tafel step). In the V-H mechanism, unlike the V-T mechanism, $H^*$ tends to bind to one of the $H^+$ in the electrolyte after the Volmer reaction, subsequently forming $H_2$ gas (Heyrovsky step).

As shown in Figure 6, in order to further understand the HER mechanism of TM@MoSi$_2$N$_4$ catalyst, we further studied the H$_2$ generation process. The results show that the energy barrier of Tafel reaction is 3.89 eV and that of Heyrovsky reaction is 1.55 eV. By comparing the barriers, it can be seen that Ni@MoSi$_2$N$_4$ prefers the Heyrovsky reaction to the Tafel reaction in the HER process.

## 4. CONCLUSIONS

In conclusion, through theoretical calculation, we screened six kinds of TM atoms (including Sc, Ti, V, Fe, Co and Ni) anchored on the MoSi$_2$N$_4$ base plane as the active sites for HER. We found that almost all TM@MoSi2N4 are well stabilized and the TM can be stabilized to anchor to the surface. More interestingly, ΔG$_{H*}$ values of most systems are significantly lower than those of the original MoSi$_2$N$_4$ system. In particular, V@MoSi$_2$N$_4$ and Ni@MoSi$_2$N$_4$ calculated ΔG$_{H*}$ to be only -0.07 eV and 0.06 eV. It indicates that TM atom anchoring is an effective method to improve HER performance of MoSi$_2$N$_4$. This is caused by the introduction of impurity levels in the band gap by TM atoms, which can increase the HER activity by decreasing the band gap to increase the conductivity. Our results provide a viable pathway to trigger HER catalytic activity of MoSi$_2$N$_4$ monolayer, thus expanding its application.

**Notes**

The authors declare no competing financial interest.

**Acknowledgements**

This work was supported by the National Natural Science Foundation of China (grant no. 12004295). P. L. thanks China's Postdoctoral Science Foundation funded project (grant no. 2022M722547).

References

1. M. K. Debe, Electrocatalyst Approaches and Challenges for Automotive Fuel Cells. Nature 486 (2012) 43−51.
2. Y. Zheng, Y. Jiao, M. Jaroniec, S. Z. Qiao, Advancing the Electrochemistry of the

Hydrogen- Evolution Reaction through Combining Experiment. Angew. Chem., Int. Ed. 54 (2015) 52−65.

3. P. C. K. Vesborg, B. Seger, I. Chorkendorff, Recent Development in Hydrogen Evolution Reaction Catalysts and Their Practical Implementation. J. Phys. Chem. Lett. 6 (2015) 951−957.

4. N.S. Lewis, D.G. Nocera, Powering the planet: Chemical challenges in solar energy utilization, Proc. Natl. Acad. Sci. U.S.A. 103 (2006) 15729–15735.

5. M. Zhang, L. Dai, Carbon nanomaterials as metal-free catalysts in next generation fuel cells, Nano Energy 1 (2012) 514–517.

6. J. A. Turner, Sustainable Hydrogen Production. Science 305 (2004) 972−974.

7. Z. Chen, P. Li, R. Anderson, X. Wang, O. K. Farha, et al. Balancing volumetric and gravimetric uptake in highly porous materials for clean energy. Science 368 (2020) 297−303.

8. S. S. Veroneau, D. G. Nocera, Continuous electrochemical water splitting from natural water sources via forward osmosis. Proc. Natl. Acad. Sci. U. S. A. 118 (2021) e2024855118.

9. J. Greeley, T.F. Jaramillo, J. Bonde, I. Chorkendorff, J.K. Nørskov, Computational high-throughput screening of electrocatalytic materials for hydrogen evolution, Nature Mater. 5 (2006) 909–913.

10. Q. Shao, F. Li, Y. Chen, X. Huang, The advanced designs of high-performance platinum-based electrocatalysts: recent progresses and challenges, Adv. Mater. Interfaces 5 (2018) 1800486.

11. J.N. Tiwari, S. Sultan, C.W. Myung, T. Yoon, N. Li, M. Ha, A.M. Harzandi, H. J. Park, D.Y. Kim, S.S. Chandrasekaran, W.G. Lee, V. Vij, H. Kang, T.J. Shin, H. S. Shin, G. Lee, Z. Lee, K.S. Kim, Multicomponent electrocatalyst with ultralow Pt loading and high hydrogen evolution activity, Nat. Energy 3 (2018) 773–782.

12. H. Jin, S. Sultan, M. Ha, J.N. Tiwari, M.G. Kim, K.S. Kim, Simple and scalable mechanochemical synthesis of noble metal catalysts with single atoms toward highly efficient hydrogen evolution, Adv. Funct. Mater. 30 (2020) 2000531.

13. X. Ping, D. Liang, Y. Wu, X. Yan, S. Zhou, D. Hu, X. Pan, P. Lu, L. Jiao, Activating a

Two-Dimensional PtSe$_2$ Basal Plane for the Hydrogen Evolution Reaction through the Simultaneous Generation of Atomic Vacancies and Pt Clusters, Nano Lett. 21 (2021) 3857–3863.

14. J. Liu, X. Fan, C.Q. Sun, W. Zhu, DFT study on bimetallic Pt/Cu(1 1 1) as efficient catalyst for H2 dissociation, Appl. Surf. Sci. 441 (2018) 23–28.

15. M. Bao, I.S. Amiinu, T. Peng, W. Li, S. Liu, Z. Wang, Z. Pu, D. He, Y. Xiong, S. Mu, Surface evolution of PtCu alloy shell over Pd nanocrystals leads to superior hydrogen evolution and oxygen reduction reactions. ACS Energy Lett. 3 (2018) 940–945.

16. Z. Cao, Q. Chen, J. Zhang, H. Li, Y. Jiang, S. Shen, G. Fu, B.A. Lu, Z Xie., L. Zheng, Platinum-nickel alloy excavated nano-multipods with hexagonal close-packed structure and superior activity towards hydrogen evolution reaction. Nat Commun. 8 (2017) 1–7.

17. B. Cui, B. Hu, J. Liu, M. Wang, Y. Song, K. Tian, Z. Zhang, L. He, Solution Plasma-assisted Bimetallic Oxide Alloy Nanoparticles of Pt and Pd Embedded within Two-dimensional Ti$_3$C$_2$T$_x$ Nanosheets as Highly Active Electrocatalysts for Overall Water-splitting. ACS Appl. Mater. Interfaces 10 (2018) 23858−23873.

18. W. Xun, Y. J. Wang, R. L. Fan, Q. Q. Mu, S. Ju, Y. Peng, M. R. Shen, Activating the MoS$_2$ Basal Plane toward Enhanced Solar Hydrogen Generation via in Situ Photoelectrochemical Control. ACS Energy Lett. 6 (2021) 267−276.

19. Z. Q. Xie, Y. Wu, Y. Zhao, M. Y. Wei, Q. S. Jiang, X. Yang, W. Xun, Activating MoS$_2$ Basal Plane via Non-noble Metal Doping For Enhanced Hydrogen Production. ChemistrySelect 8 (2023) e202204608.

20. W. Xun, Y. Zhao, M. Y. Wei, X. Yang, S. Q. Cao, R. G. Ma, Q. S. Jiang. Activating and optimizing MoS$_2$ basal-plane via spontaneous oxidation for enhanced photocatalytic hydrogen generation. Materials Today Communications 36 (2023) 106609.

21. J. Lee, S. Kang, K. Yim, K.Y. Kim, H.W. Jang, Y. Kang, S. Han, Hydrogen evolution reaction at anion vacancy of two-dimensional transition-metal dichalcogenides: ab initio computational screening, J. Phys. Chem. Lett. 9 (2018) 2049–2055.

22. Q. P. Lu, Y. F. Yu, Q, L. Ma, B. Chen, H. Zhang, 2D Transition-Metal-Dichalcogenide-Nanosheet-Based Composites for Photocatalytic and Electrocatalytic Hydrogen Evolution Reactions. Adv. Mater. 28 (2016) 1917–1933.

23. Q. Fu, J. Han, X. J. Wang, P. Xu, T. Yao, J. Zhong, W. W. Zhong, S. W. Liu, T. L. Gao, Z. H. Zhang, L. L. Xu, B. Song, 2D Transition Metal Dichalcogenides: Design, Modulation, and Challenges in Electrocatalysis. Adv. Mater. 2020, 1907818.

24. M. J. Liu, M. S. Hybertsen, Q. Wu, A Physical Model for Understanding the Activation of $MoS_2$ Basal-Plane Sulfur Atoms for the Hydrogen Evolution Reaction. Angew. Chem. Int. Ed. 59 (2020) 2–9.

25. G. P. Gao, A. P. O'Mullane, A. J. Du, 2D MXenes: A New Family of Promising Catalysts for the Hydrogen Evolution Reaction. ACS Catal. 7 (2017) 494−500.

26. H. C. Yang, Y. D. Ma, X. S. Lv, B. B. Huang, Y. Dai, Prediction of intrinsic electrocatalytic activity for hydrogen evolution reaction in $Ti_4X_3$ (X = C, N). Journal of Catalysis 387 (2020) 12–16.

27. M. Zubair, M. M. Ul Hassan, M. T.Mehran, M. M. Baig, S. Hussain, F. Shahzad, 2D MXenes and their heterostructures for HER, OER and overall water splitting: A review. International Journal of Hydrogen Energy 47 (2022) 2794-2818.

28. S. S. Bai, M. Q. Yang, J. Z. Jiang, X. M. He, J. Zou, Z. G. Xiong, G. D. Liao, S. Liu, Recent advances of MXenes as electrocatalysts for hydrogen evolution reaction. npj 2D Materials and Applications volume 5 (2021) 78.

29. Y. F. Zhao, J. Q. Zhang, X. Guo, X. J. Cao, S. J. Wang, H. Liu, G. X. Wang, Engineering strategies and active site identification of MXene-based catalysts for electrochemical conversion reactions. Chem. Soc. Rev. 52 (2023) 3215-3264.

30. D. H. Deng, K. S. Novoselov, Q. Fu, N. F. Zheng, Z. Q. Tian, X. H. Bao, Catalysis with two-dimensional materials and their heterostructures, Nat. Nanotechnol. 11 (2016) 218–230.

31. K. Hu, T. Ohto, Y. Nagata, M. Wakisaka, Y. Aoki, J. Fujita, Y. Ito, Catalytic activity of graphene-covered non-noble metals governed by proton penetration in electrochemical hydrogen evolution reaction. Nat. Commun. 12 (2021) 203.

32. D. Wei, L. Chen, L. Tian, S. Ramakrishna, D. Ji, Hierarchically Structured CoNiP/CoNi Nanoparticle/Graphene/Carbon Foams as Effective Bifunctional Electrocatalysts for HER and OER. Ind. Eng. Chem. Res. 62 (2023) 4987–4994.

33. Q. Yang, H. Liu, P. Yuan, Y. Jia, L. Zhuang, H. Zhang, X. Yan, G. Liu, Y. Zhao, J. Liu, S.

Wei, L. Song, Q. Wu, B. Ge, L. Zhang, K. Wang, X. Wang, C. Chang, X. Yao, Single Carbon Vacancy Traps Atomic Platinum for Hydrogen Evolution Catalysis. J. Am. Chem. Soc. 144 (2022) 2171–2178.

34. Q. Yang, H. X. Liu, P. Yuan, Y. Jia, L. Z. Zhuang, H. W. Zhang, X. C. Yan, G. H. Liu, Y. F. Zhao, J. Z. Liu, S. Q. Wei, L. Song, Q. L. Wu, B. Q. Ge, L. Z. Zhang, K. Wang, X. Wang, C. R. Chang, X. D. Yao, Single Carbon Vacancy Traps Atomic Platinum for Hydrogen Evolution Catalysis. J. Am. Chem. Soc. 144 (2022) 2171–2178.

35. Y. Xue, B. Huang, Y. Yi, Y. Guo, Z. Zuo, Y. Li, Z. Jia, H. Liu, Y. Li, Anchoring zero valence single atoms of nickel and iron on graphdiyne for hydrogen evolution. Nat. Commun. 9 (2018) 1460.

36. W. H. Lai, L. F. Zhang, W. B. Hua, S. Indris, Z. C. Yan, Z. Hu, B. Zhang, Y. Liu, L. Wang, M. Liu, R. Liu, Y. X. Wang, J. Z. Wang, Z. Hu, H. K. Liu, S. L. Chou, S. X. Dou, General π-Electron-Assisted Strategy for Ir, Pt, Ru, Pd, Fe, Ni Single-Atom Electrocatalysts with Bifunctional Active Sites for Highly Efficient Water Splittin. Angew. Chem. Int. Ed. Engl. 58 (2019) 11868– 11873,

37. Y. D. Peng, K. Ma, T. Z. Xie, J. Q. Du, L. R. Zheng, F. B. Zhang, X. B. Fan, W. C. Peng, J. Y. Ji, Y. Li, Tunable Pt–Ni Interaction Induced Construction of Disparate Atomically Dispersed Pt Sites for Acidic Hydrogen Evolution. ACS Applied Materials & Interfaces, 15 (2023) 27089-27098.

38. G. L. Liu, A. W. Robertson, M. M. Li, W. C. H. Kuo, M. T. Darby, M. H. Muhieddine, Y. Lin, K. Suenaga, M. Stamatakis, J. H. Warner, S. C. E. Tsang, $MoS_2$ monolayer catalyst doped with isolated Co atoms for the hydrodeoxygenation reaction. Nature chemistry 9 (2017) 2740.

39. T. H. M. Lau, X. W. Lu, J. Kulhavy, S. Wu, L. L. Lu, T. S. Wu, R. Kato, J. S. Foord, Y. Soo, K. Suenagad, S. C. E. Tsang, Transition metal atom doping of the basal plane of $MoS_2$ monolayer nanosheets for electrochemical hydrogen evolution. Chem. Sci. 9 (2018) 4769.

40. H. S. Li, S. S. Wang, H. Sawada, G.G. D. Han, T. Samuels, C. S. Allen, A. I. Kirkland, J. C. Grossman, J. H. Warner, Atomic Structure and Dynamics of Single Platinum Atom Interactions with Monolayer $MoS_2$. ACS Nano 11 (2017) 3392−3403.

41. S. Tian, Z. Wang, W. Gong, W. Chen, Q. Feng, Q. Xu, C. Chen, C. Chen, Q. Peng, L. Gu, H. Zhao, P. Hu, D. Wang, Y. Li, Temperature-controlled selectivity of hydrogenation and hydrodeoxygenation in the conversion of biomass molecule by the $Ru_1$/mpg-$C_3N_4$ catalyst. J. Am. Chem. Soc. 140 (2018) 11161–11164.

42. F. Yu, T. Huo, Q. Deng, G. Wang, Y. Xia, H. Li, W. Hou, Single-atom cobalt-hydroxyl modification of polymeric carbon nitride for highly enhanced photocatalytic water oxidation: ball milling increased single atom loading. Chem. Sci. 13 (2022) 754.

43. Mingming Zhang, Cui Lai, Bisheng Li, Shiyu Liu, Danlian Huang, Fuhang Xu, Xigui Liu, Lei Qin, Yukui Fu, Ling Li, Huan Yi, and Liang Chen, MXenes as Superexcellent Support for Confining Single Atom: Properties, Synthesis, and Electrocatalytic Applications. Small 17 (2021) 2007113.

44. J. L. Suter, R. C. Sinclair, P. V. Coveney, Principles Governing Control of Aggregation and Dispersion of Graphene and Graphene Oxide in Polymer Melts. Adv. Mater. 32 (2020) 200321.

45. J. Pető, T. Ollár, P. Vancsó, Z. I. Popov, G. Z. Magda1, G. Dobrik1, C. Hwang, P. B. Sorokin, L. Tapasztó, Spontaneous doping of the basal plane of $MoS_2$ single layers through oxygen substitution under ambient conditions. Nature chem. 10 (2018) 1246–1251.

46. R. A. Soomro, P. Zhang, B. M. Fan, Y. Wei, B. Xu, Progression in the Oxidation Stability of MXenes, Nano-Micro Lett. 15 (2023)108.

47. Y. L. Hong, Z. Liu, L. Wang, T. Zhou, W. Ma, C. Xu, S. Feng, L. Chen, M. L. Chen, D. M. Sun, X. Q. Chen, H. M. Cheng, W. C. Ren, Chemical vapor deposition of layered two-dimensional $MoSi_2N_4$ materials. Science 369 (2020) 670−674.

48. P. Li, X. Yang, Q. S. Jiang, Y. Z. Wu, W. Xun, Built-in electric field and strain tunable valley-related multiple topological phase transitions in $VSi_XN_4$ (X = C, Si, Ge, Sn, Pb) monolayers. Phys. Rev. Mater. 7 (2023) 064002.

49. B. Mortazavi, B. Javvaji, F. Shojaei, T. Rabczuk, A. V. Shapeev, X. Zhuang, Exceptional Piezoelectricity, High Thermal Conductivity and Stiffness and Promising Photocatalysis in Two-Dimensional $MoSi_2N_4$ Family Confirmed by First-Principles. Nano Energy 82 (2021) 105716.

50. B. Li, J. Geng, H. Ai, Y. Kong, H. Bai, K. H. Lo, K. W. Ng, Y. Kawazoe, H. Pan, Design of 2D Materials-$MSi_2C_xN_{4-x}$ (M = Cr, Mo, and W; X = 1 and 2)-with Tunable Electronic and Magnetic Properties. Nanoscale 13 (2021) 8038−8048.

51. S. Li, W. Wu, X. Feng, S. Guan, W. Feng, Y. Yao, S. A. Yang, Valley-Dependent Properties of Monolayer $MoSi_2N_4$, $WSi_2N_4$, and $MoSi_2As_4$. Phys. Rev. B: Condens. Matter Mater. Phys. 102 (2020) 113102.

52. C. Yang, Z. Song, X. Sun, J. Lu, Valley Pseudospin in Monolayer $MoSi_2N_4$ and $MoSi_2As_4$. Phys. Rev. B: Condens. Matter Mater. Phys. 103 (2021) 035308.

53. J. Chen, Q. Tang, The Versatile Electronic, Magnetic and Photo-Electro Catalytic Activity of a New 2D $MA_2Z_4$ Family. Chem. Eur. J. 27 (2021) 9925−9933.

54. Y. Chen, S. Tian, Q. Tang, First-Principles Studies on Electrocatalytic Activity of Novel Two-Dimensional $MA_2Z_4$ Monolayers toward Oxygen Reduction Reaction. J. Phys. Chem. C, 125 (2021) 22581−22590.

55. Y. Yu, J. Zhou, Z. Guo, Z. Sun, Novel Two-Dimensional Janus $MoSiGeN_4$ and $WSiGeN_4$ as Highly Efficient Photocatalysts for Spontaneous Overall Water Splitting. ACS Appl. Mater. Interfaces 13 (2021) 28090−28097.

56. W. Xun, X. Yang, Q. S. Jiang, M. J. Wang, Y. Z. Wu, P. Li, Single-Atom-Anchored Two-Dimensional $MoSi_2N_4$ Monolayers for Efficient Electroreduction of $CO_2$ to Formic Acid and Methane. ACS Appl. Energy Mater. 6 (2023) 3236−3243.

57. Y. Zang, Q. Wu, W. Du, Y. Dai, B. Huang, Y. Ma, Activating electrocatalytic hydrogen evolution performance of two-dimensional $MSi_2N_4$(M = Mo, W): A theoretical prediction. Phys. Rev. Mater. 5 (2021) 045801.

58. W. Shi, G. Yin, S. Yu, T. Hu, X. Wang, Z. Wang, Atomic precision tailoring of two-dimensional $MoSi_2N_4$ as electrocatalyst for hydrogen evolution reaction. J. Mater. Sci. 57 (2022) 18535–18548.

59. C. L. Zhang, Y. Z. Yuan, B. N. Jia, F. Wei, X. H. Zhang, G. Wu, L. Li, C. C. Chen, Z. Q. Zhao, F Chen, J. B. Hao, P. F. Lu, Defect engineered Janus $MoSiGeN_4$ as highly efficient electrocatalyst for hydrogen evolution reaction. Appl. Surf. Sci. 622 (2023) 156894.

60. Y. Y. Liu, Y. J. Ji, Y. Y. Li, Multilevel Theoretical Screening of Novel Two-Dimensional

MA$_2$Z$_4$ Family for Hydrogen Evolution. J. Phys. Chem. Lett. 12 (2021) 9149−9154.

61. J. N. Zheng, X. Sun, J. X. Hu, S. B. Wang, Z. H. Yao, S. W. Deng, X. Pan, Z. Y. Pan, J. G. Wang, Symbolic Transformer Accelerating Machine Learning Screening of Hydrogen and Deuterium Evolution Reaction Catalysts in MA$_2$Z$_4$ Materials. ACS Appl. Mater. Interfaces 13 (2021) 50878−50891.

62. W. W. Qian, Z. Chen, J. F. Zhang, L. C. Yin, Monolayer MoSi$_2$N$_{4-x}$ as promising electrocatalyst for hydrogen evolution reaction: A DFT prediction. J. Mater. Sci. Technol. 99 (2022) 215–222.

63. C. W. Xiao, R. J. Sa, Z. T. Cui, S. S. Gao, W. Du, X. Q. Sun, X. T. Zhang, Q. H. Li, Z. J. Ma, Enhancing the hydrogen evolution reaction by non-precious transition metal (Non-metal) atom doping in defective MoSi$_2$N$_4$ monolayer. Appl. Surf. Sci. 622 (2023) 156894.

64. M. R. Sahoo, A. Ray, N. Singh, Theoretical Insights into the Hydrogen Evolution Reaction on VGe$_2$N$_4$ and NbGe$_2$N$_4$ Monolayers. ACS Omega 7 (2022) 7837−7844.

65. G. Kresse, J. Furthmüller, Efficiency of ab-initio total energy calculations for metals and semiconductors using a plane-wave basis set. Comput. Mater. Sci. 6 (1996) 15−50.

66. G. Kresse, J. Furthmüller, Efficient iterative schemes for ab initio total-energy calculations using a plane-wave basis set. Phys. Rev. B 54 (1996) 11169−11186.

67. J. Wellendorff, K. T. Lundgaard, A. Møgelhøj, V. Petzold, D. D. Landis, J. K. Nørskov, T. Bligaard, K. W. Jacobsen, Density functionals for surface science: Exchange-correlation model development with Bayesian error estimation. Phys. Rev. B 85 (2012) 235149.

68. J. P. Perdew, M. Ernzerhof, K. Burke, Rationale for mixing exact exchange with density functional approximations. J. Chem. Phys. 105 (1996) 9982−9985.

69. J. P. Perdew, K. Burke, M. Ernzerhof, Generalized gradient approximation made simple. Phys. Rev. Lett. 77 (1996) 3865−3868.

70. M. Tuckerman, K. Laasonen, M. Sprik, M. J. Parrinello, Ab initio molecular dynamics simulation of the solvation and transport of hydronium and hydroxyl ions in water, J. Chem. Phys. 103 (1995) 150–161.

71. K. Mathew, R. Sundararaman, K. Letchworth-Weaver, T. Arias, R. G. Hennig, Implicit solvation model for density-functional study of nanocrystal surfaces and reaction

pathways, J. Chem. Phys. 140 (2014), 084106.

72. J. Rossmeisl, Z. W. Qu, H. Zhu, G. J. Kroes, J. K Nørskov, Electrolysis of water on oxide surfaces, J. Electroanal. Chem. 607 (2007) 83–89.

73. J. K. Nørskov, T. Bligaard, A. Logadottir, J. Kitchin, J. G. Chen, S. Pandelov, U. Stimming, Trends in the exchange current for hydrogen evolution, J. Electrochem. Soc. 152 (2005) J23.

74. Y. Li, Z. Zhou, G. Yu, W. Chen, Z. Chen, Catalytic oxidation on iron-embedded graphene: computational quest for low-cost nanocatalysts. J. Phys. Chem. C. 114 (2010) 6250–6254.

75. F. Li, Y. Li, X. C. Zeng, Z. Chen, Exploration of high-performance single atom catalysts on support $M_1/FeO_x$ for CO oxidation via computational study. ACS Catal. 5, 544–552 (2014).

76. P. Li, J. Zhu, A.D. Handoko, R. Zhang, H. Wang, D. Legut, X. Wen, Z. Fu, Z.W. Seh, Q. Zhang, High-throughput theoretical optimization of the hydrogen evolution reaction on MXenes by transition metal modification, J. Mater. Chem. A 6 (2018) 4271–4278.

77. X. Zhao, Y.Yang, Y. J. Hu, G. Wang, D. G. Wang, Y. F. Wei, S. X. Zhou, J. S. Bi, W. J. Xiao, X. F. Liu, Theoretical calculation of hydrogen evolution reaction in two-dimensional $As_2X_3$(X=S, Se, Te) doped with transition metal atoms. Appl. Surf. Sci. 616 (2023) 156475.